\documentstyle[12pt,twoside]{article}

\def\d{\delta}

\def\k{\kappa}

\def\m{\mu}
\def\n{\nu}
\def\o{\omega}

\def\s{\sigma}

\def\D{\Delta}

\def\bo{{\raise.15ex\hbox{\large$\Box$}}}

\def\TH{{\raise.2ex\hbox{$\displaystyle \bigodot$}\mskip-4.7mu \llap H \;}}
\def\face{{\raise.2ex\hbox{$\displaystyle \bigodot$}\mskip-2.2mu \llap {$\ddot
        \smile$}}}

\def\Hat#1{\rlap{\kern.10em$\widehat{\phantom G}$}#1}
\def\HAt#1{\rlap{\kern.05em$\widehat{\phantom G}$}#1}

\def\capp#1{\rlap{\kern.1em$\widehat{\phantom{G\vrule height.8em}}$}#1{}}
\def\Capp#1{\rlap{\kern.05em$\widehat{\phantom{G\vrule height.8em}}$}#1{}}

\def\leftrightarrowfill{$\mathsurround=0pt \mathord\leftarrow \mkern-6mu
        \cleaders\hbox{$\mkern-2mu \mathord- \mkern-2mu$}\hfill
        \mkern-6mu \mathord\rightarrow$}
\def\overleftrightarrow#1{\vbox{\ialign{##\crcr
        \leftrightarrowfill\crcr\noalign{\kern-1pt\nointerlineskip}
        $\hfil\displaystyle{#1}\hfil$\crcr}}}

\def\frac#1#2{{\textstyle{#1\over\vphantom2\smash{\raise.20ex
        \hbox{$\scriptstyle{#2}$}}}}}

\def\sfrac#1#2{{\vphantom1\smash{\lower.5ex\hbox{\small$#1$}}\over
        \vphantom1\smash{\raise.4ex\hbox{\small$#2$}}}}
\def\bfrac#1#2{{\vphantom1\smash{\lower.5ex\hbox{$#1$}}\over
        \vphantom1\smash{\raise.3ex\hbox{$#2$}}}}
\def\afrac#1#2{{\vphantom1\smash{\lower.5ex\hbox{$#1$}}\over#2}}

\catcode`@=11
\def\underline#1{\relax\ifmmode\@@underline#1\else
        $\@@underline{\hbox{#1}}$\relax\fi}
\catcode`@=12

\def\nis{\nointerlineskip}
\def\Abar{\vbox{\nis\moveright.33em\vbox{
        \hrule width.35em height.04em}\nis\kern.05em\hbox{$A$}}{}}
\def\Dbar{\vbox{\nis\moveright.20em\vbox{
        \hrule width.50em height.04em}\nis\kern.05em\hbox{$D$}}{}}
\def\Gbar{\vbox{\nis\moveright.20em\vbox{
        \hrule width.50em height.04em}\nis\kern.05em\hbox{$G$}}{}}
\def\mbar{\vbox{\nis\moveright.15em\vbox{
        \hrule width.60em height.04em}\nis\kern.05em\hbox{$m$}}{}}
\def\Rbar{\vbox{\nis\moveright.20em\vbox{
        \hrule width.50em height.04em}\nis\kern.05em\hbox{$R$}}{}}
\def\Vbar{\vbox{\nis\moveright.05em\vbox{
        \hrule width.60em height.04em}\nis\kern.05em\hbox{$V$}}{}}
\def\Xbar{\vbox{\nis\moveright.20em\vbox{
        \hrule width.60em height.04em}\nis\kern.05em\hbox{$X$}}{}}
\def\thetabar{\vbox{\nis\moveright.15em\vbox{
        \hrule width.30em height.04em}\nis\kern.05em\hbox{$\theta$}}{}}
\def\Lambdabar{\vbox{\nis\moveright.25em\vbox{
        \hrule width.35em height.04em}\nis\kern.05em\hbox{${\mit\Lambda}$}}{}}
\def\Sigmabar{\vbox{\nis\moveright.25em\vbox{
        \hrule width.50em height.04em}\nis\kern.05em\hbox{${\mit\Sigma}$}}{}}
\def\phibar{\vbox{\nis\moveright.18em\vbox{
        \hrule width.40em height.04em}\nis\kern.05em\hbox{$\phi$}}{}}
\def\chibar{\vbox{\nis\moveright.12em\vbox{
        \hrule width.40em height.04em}\nis\kern.05em\hbox{$\chi$}}{}}
\def\psibar{\vbox{\nis\moveright.23em\vbox{
        \hrule width.40em height.04em}\nis\kern.05em\hbox{$\psi$}}{}}
\def\debar{\vbox{\nis\moveright.18em\vbox{
        \hrule width.35em height.04em}\nis\kern.05em\hbox{$\partial$}}{}}
\def\delbar{\vbox{\nis\moveright.10em\vbox{
        \hrule width.63em height.04em}\nis\kern.05em\hbox{$\nabla$}}{}}

\newskip\humongous \humongous=0pt plus 1000pt minus 1000pt

\newif\ifdtup

\def\begintitle#1#2#3#4
        {\begin{titlepage}
         \centerline{#1 \hfill MdDP-PP-89-#2}
         \begin{center}\vglue .7in
         {\large\bf #3}\\[.7in]
         {\bf #4}\\
         {\it Department of Physics and Astronomy}\\
         {\it University of Maryland, College Park, MD 20742}\\[.7in]
         {\bf ABSTRACT}\\
         \end{center}
         \begin{quotation}}
\def\endtitle
         {\end{quotation}
          \end{titlepage}
          \newpage}

\newcommand{\comment}[1]{}

\newcommand{\modestsections}{
    \let\section=\subsection
        \renewcommand{\thesubsection}{\arabic{subsection}}
    \let\subsection=\subsubsection  
    \let\subsubsection=\paragraph
    \let\paragraph=\subparagraph
    \renewcommand{\subparagraph}[1]{\paragraph{##1}
        \typeout{You've used the subparagraph command, which is the same as
                 the paragraph command since you're using modest sections.}}}

\newtheorem{theorem}{Theorem}[section]

\newcommand{\pfhead}{\noindent {\bf Proof:} }

\def\squarebox#1{\hbox to #1{\hfill\vbox to #1{\vfill}}}
\newcommand{\qedbox}{\vbox{\hrule\hbox{\vrule\squarebox{.667em}\vrule}\hrule}}
\newcommand{\qed}{\nopagebreak\mbox{}\hfill\qedbox\smallskip}
\begin{document}

\title{No Parity Violation without R- Parity Violation$^{\dagger}$}

\author{Ravi Kuchimanchi and R.N. Mohapatra  \\
\vspace{.1in} \\
Department of Physics, University of Maryland\\
College Park, Maryland 20742\\
UMP-PP-93-209\\
June, 1993 \\
hep-ph@xxx.lanl.gov bulletin board number: hep-ph/9306290}

\date{}
\maketitle

\begin{abstract}
In a class of minimal supersymmetric left-right models (SUSYLR)
of weak interactions
where R-parity is automatically conserved, we show that spontaneous breakdown
of parity cannot occur without spontaneous breakdown of R-parity.  This
intriguing result
connects two physically different scales in supersymmetric models.
\end{abstract}
\vskip0.2in
Note: Revised to include Table.tex, ie Table 1, that we had forgotten in an
earlier mailing.
\vspace*{2.0in}
\vskip0.5in
\noindent$^\dagger$ Work supported by a grant from the National Science
Foundation.
\newpage
\section{Introduction}

An awkward aspect of the minimal supersymmetric standard model (MSSM)
is the presence of R-parity\footnote{A particle of Baryon number $B$,
Lepton number $L$ and spin $S$ has R-parity $(-1)^{3B + L + 2 S}.$
Thus all the familiar particles like the
quarks, leptons and Higgs boson are R-parity even and their
superpartners are R-parity odd.}
violating terms in the superpotential~$\left [ 1 \right ]$,
which are allowed by gauge invariance.  These terms lead to lepton and baryon
number violation and their strengths are therefore severely
limited by phenomenological~$\left [ 1 \right ]$ and cosmological constraints
In fact unless the strength of the baryon number violating
term is less than $10^{- 13}$, it will lead to
contradiction with present lower limits on the lifetime of the proton.
It would therefore be more appealing to have a supersymmetric theory
where R - parity conservation is automatic~$\left [ 3, 4 \right ]$.  It was
noted some time ago~$\left [ 3 \right ]$ that if the gauge symmetry of the
supersymmetric model is extended to $SU(2)_{L} \times U(1)_{I_{3R}}
\times U(1)_{B - L}$ or $SU(2)_{L} \times SU(2)_{R}
\times U(1)_{B - L}$,
the theory becomes {\it automatically} R - parity conserving.
One can then entertain the
possibility of spontaneous R - parity violation~$\left [ 5, 6 \right ]$.
There are
two distinct differences between spontaneous and explicit
R -parity violation.  The first is that the catastrophic baryon number
violating terms are automatically absent from the Lagrangian,
in the spontaneous R -Parity violation case,so
that proton stability is guaranteed; and the second is that,
in this case,  above a certain
temperature, R - parity is restored so that the
picture of the early universe is very different;
for instance, any pre-existing lepton
or baryon asymmetry of the
universe need not be erased.  Furthermore, the strength of R - parity
violating terms is not anymore arbitrary but is connected to the scale
of R-parity breaking.

In this paper,
we show that in the minimal supersymmetric left - right (SUSYLR)
model with the see-saw mechanism~$\left [ 7 \right ]$,
the spontaneous breaking of Parity $\left [ 8 \right ]$
and that of
R- parity are intimately linked.  By a detailed analysis of the Higgs potential
of the model, we show that if
R - parity remains unbroken so does the parity
symmetry of the model, even after one-loop radiative corrections are taken into
account.  Therefore, a necessary condition for parity violation to occur
is the
existence of spontaneous breakdown of R - parity.  This is in our opinion an
interesting result since it connects the breaking of two different symmetries
and links their breaking scales.
Since, in this class of models, the scale of parity breaking
is connected to the neutrino mass $\left [ 9 \right ]$
via the see-saw mechanism,
the R-parity breaking scale and the strength of the induced R-parity breaking
interactions get related to the neutrino masses.
Some aspects of symmetry breaking for these
models were studied earlier in reference~$\left [ 10 \right ]$
but this connection between
breaking R - parity and Parity was not noticed in this paper.

The rest of the paper is organized as follows - in section~\ref{sec:higgs}
we write down the most general tree level Higgs potential for the minimal
SUSYLR model; in section~\ref{sec:min} we derive the inequality constraints
between the
parameters of the tree level Higgs potential by requiring that it be
bounded from below and use these constraints to prove that there can
be no parity violation without R-parity violation
for the tree level potential; in section~\ref{sec:nonmin},
we show that even in certain non-minimal SUSYLR models R - parity must be
violated or else either electric charge is not conserved
by the ground state of the theory or Parity symmetry is
unbroken; in section~\ref{sec:rpar} we prove that if R - Parity is violated
spontaneously,
then in the minimal SUSYLR model itself,
the ground state can violate parity
and conserve electric
charge and
in section~\ref{sec:rad} we use the freedom of choosing
a convenient renormalization scale to argue that in the minimal SUSYLR
model radiative corrections do not change the tree level result that
there can be no parity violation without R-parity violation, since
otherwise electric
charge is not conserved.
\section{The Higgs potential for the SUSY Left - Right Model}
\label{sec:higgs}
The matter content of the minimal model is given in Table~1.
\vskip0.2in

 $~~~~~~~~~~~~~~~~~~~~~$
\begin{tabular}{|c|c|} \hline
Matter Superfield & $SU(2)_{L} \times SU(2)_{R} \times U(1)_{B-L}$ \\
                   & Quantum Number \\ \hline
$Q$                  & $(2, 1, {1 \over 3})$\\
$Q^{c}$        & $(1, 2, - {1\over 3})$  \\
$L$             & $(2, 1, -1)$ \\
$L^{c}$        & $(1, 2, +1)$ \\ \hline
Higgs Fields      &  \\ \hline
$\D$            & $(3, 1, +2)$  \\
$\D^{c}$        & $(1, 3, -2)$ \\
$\overline{\D}$ & $(3, 1, -2)$ \\
$\overline{\D}^{c}$ & $(1, 3, +2)$ \\
$\Phi$      & $(2, 2, 0)$ \\ \hline
\end{tabular}

\vskip0.2in

 $~~~~~~~~~~~~~~~~~~~~$Table 1: The matter content of minimal SUSYLR model.

\vskip0.2in

In this paper the doublets are represented by $2 \times 1$ column vectors,
the triplets by $2 \times 2$ traceless
complex matrices and the bidoublet $\Phi$
is represented by $2 \times 2$ complex matrix. Also note that\footnote{In
this paper $\tau_{1}, \tau_{2}, \tau_{3}$ stand for the Pauli matrices.}
, for example,
under the $SU(2)_{L} \times SU(2)_{R}$ part of the
gauge transformation, $L \rightarrow U_{L}L$, $\D \rightarrow
U_{L} \D U_{L}^{\dagger}$, $\tau_{2} L^{c} \rightarrow U_{R}\left(\tau_{2}
L^{c}\right )$, $\tau_{2} \D^{c} \tau_{2} \rightarrow U_{R} \left ( \tau_{2}
\D^{c}
\tau_{2} \right ) U_{R}^{\dagger}$
and $\Phi \rightarrow U_{L} \Phi U_{R}^{\dagger}$
where $U_{L}$ and $U_{R}$ are the $SU(2)_{L}$ and $SU(2)_{R}$
group transformations (similarly
for $\overline{\D}$ and $\overline{\D}^{c}$).
The gauge-invariant superpotential $\left [ 10, 11 \right ]$ for
the model is given by\footnote{Note that we are imposing an
exact discrete parity
symmetry in the usual way}:
\begin{eqnarray}
\label{eq:super}
W = & ~& h_{q} Q^{T} \tau_{2} \Phi \tau_{2} Q^{c} + h L^{T} \tau_{2}
\Phi\tau_{2} L^{c} \nonumber \\
& ~& + i f \left ( L^{T} \tau_{2} \D L + L^{cT} \tau_{2} \D^{c} L^{c} \right)
\nonumber \\
& ~& + M ~Tr\left ( \D \overline{\D} + \D^{c} \overline{\D}^{c} \right )
 + \mu' ~Tr \left ( \tau_{2} \Phi^{T} \tau_{2} \Phi \right )
\end{eqnarray}
The most general form of the Higgs potential including soft breaking terms
(but omitting the $Q$ - terms) is
given by $\left [ 10, 11 \right ]$: (Denoting the scalar
components of the superfields by $\tilde{L},
\tilde{L}^{c}, \Phi, \D, \overline{\D}, \D^{c}$
and $\overline
{\D}^{c})$
\begin{equation}
\label{eq:hig}
V = V_{F -~terms} ~~+ V_{soft} ~~+ V_{D -~terms}
\end{equation}
where
\begin{eqnarray}
\label{eq:fter}
V_{F-~terms} = & ~& \mid h \tilde{L}^{T}
\tau_{2} \Phi \tau_{2} + 2 i f \tilde{L}^{cT} \tau_{2} \D^{c}\mid^{2}
+ \mid h \tilde{L}^{cT} \tau_{2} \Phi^{T}
\tau_{2} + 2 i f \tilde{L}^{T} \tau_{2} \D\mid^{2}
\nonumber \\
& ~& + ~ Tr \mid h \tilde{L}^{c} \tilde{L}^{T} + 2 \mu' \Phi^{T} \mid^{2}
+ ~Tr \mid i f \tilde{L} \tilde{L}^{T} \tau_{2} + M \overline{\D}\mid^{2}
\nonumber \\
& ~& + ~Tr \mid i f \tilde{L}^{c}
\tilde{L}^{cT} \tau_{2} + M \overline{\D}^{c}\mid^{2}
+ \mid M \mid^{2} Tr \left ( \D^{\dagger} \D + \D^{c\dagger}\D^{c}\right),
\end{eqnarray}
\begin{eqnarray}
\label{eq:soft}
V_{soft} = & ~& m_{l}^{2} \left (
\tilde{L}^{\dagger} \tilde{L} + \tilde{L}^{c\dagger} \tilde{L}^{c} \right )
+ \left (M_{1}^{2} - \mid M \mid^{2} \right ) Tr \left (\D^{\dagger} \D
+ \D^{c\dagger} \D^{c} \right )\nonumber \\
& ~& + \left (M_{2}^{2} - \mid M \mid^{2} \right )
Tr \left (\overline{\D}^{\dagger} \overline{\D}
+ \overline{\D}^{c\dagger} \overline{\D}^{c} \right )
+ \left (M'^{2} Tr \left ( \D \overline{\D} + \D^{c} \overline{\D}^{c}
\right ) + h. c. \right ) \nonumber \\
& ~& + \left ( M_{\phi}^{2} - 4 \mid \mu' \mid^{2} \right )
Tr \Phi^{\dagger} \Phi
+ \left ( {{\mu^{2}} \over 2} Tr \left (\tau_{2} \Phi^{T} \tau_{2} \Phi
\right ) + h. c. \right ) \nonumber \\
& ~& + \left ( i v \left (\tilde{L}^{T}
\tau_{2} \D \tilde{L} + \tilde{L}^{cT} \tau_{2} \D^{c} \tilde{L}^{c}
\right ) + e \tilde{L}^{T}
\tau_{2} \Phi \tau_{2} \tilde{L}^{c} + h.c. \right ),
\end{eqnarray}
\begin{eqnarray}
\label{eq:dterm}
V_{D - terms} = & ~& {{g^{2}} \over 8}
\sum_{m} \mid \tilde{L}^{\dagger} \tau_{m} \tilde{L}
+ Tr \left ( 2 \D^{\dagger} \tau_{m} \D + 2 \overline{\D}^{\dagger} \tau_{m}
\overline{\D} + \Phi^{\dagger} \tau_{m} \Phi \right ) \mid^{2} \nonumber \\
& ~&+ {{g^{2}} \over 8}
\sum_{m} \mid \tilde{L}^{c\dagger} \tau_{m} \tilde{L}^{c}
+ Tr \left ( 2 \D^{c\dagger} \tau_{m} \D^{c}
+ 2 \overline{\D}^{c\dagger} \tau_{m} \overline{\D}^{c}
+ \Phi \tau_{m}^{T} \Phi^{\dagger} \right ) \mid^{2} \nonumber \\
& ~&+ {{g'^{2}} \over 8} \mid
\tilde{L}^{c\dagger} \tilde{L}^{c} - \tilde{L}^{\dagger} \tilde{L} +
2 Tr \left ( \D^{\dagger} \D - \D^{c\dagger} \D^{c} - \overline{\D}^{\dagger}
\overline{\D} + \overline{\D}^{c\dagger}\overline{\D}^{c}\right ) \mid^{2}
\end{eqnarray}
Note that in equation~(\ref{eq:soft}) the coefficient of $Tr\left(
\D^{\dagger} \D + \D^{c\dagger} \D^{c}\right )$ is defined
such that
the coefficient of this term
in $V_{F -terms} + V_{soft}$ is simply $M_{1}^{2}$.
Also, in the above
equations the doublets, triplets and the bidoublets have the
following electric charge quantum numbers:
\begin{equation}
\tilde{L} = \left ( \begin{array}{c}
                    \tilde{\nu} \\
                     \tilde{e}^{-}
                   \end{array}
                 \right )
{}~~, ~~~\tilde{L}^{c} = \left ( \begin{array}{c}
                    \tilde{\nu}^{c} \\
                     \tilde{e}^{+}
                   \end{array}
                 \right )
{}~~, ~~~\D = \left ( \begin{array}{cc}
                       {{\d^{+}}\over {\surd 2}} & \d^{++}\\
                         \d^{0} & - {{\d^{+}}\over {\surd 2}}
                       \end{array}
                  \right )
{}~~, ~~~\Phi = \left ( \begin{array}{cc}
                       \phi_{1}^{0} & \phi_{2}^{+}\\
                        \phi_{1}^{-} & \phi_{2}^{0}
                          \end{array}
                             \right)
\end{equation}
and similarly for other fields.

Let us note that spontaneous R -parity violation in this model arises
when either $\left<\tilde{\nu}^{c}\right> \neq 0$
and / or $\left<\tilde{\nu}\right> \neq 0$.  Also it is
obvious that parity symmetry is spontaneously broken if
\begin{equation}
\label{eq:parbkg}
\mid\left<\D\right>\mid,
\mid\left<\overline{\D}\right>\mid
{}~~<<~~~~\mid\left<\D^{c}\right>\mid ~or~
\mid\left<\overline{\D}^{c}\right>\mid
\end{equation}
We will show that if $\left<\tilde{\nu}\right>$
and $\left<\tilde{\nu}^{c}\right>$ both have zero vacuum
expectation value, the inequality~(\ref{eq:parbkg}) cannot be satisfied
and in fact the ground state of the theory corresponds to
$\left<\D\right> = \left<\overline{\D}\right> =
\left<\D^{c}\right> = \left<\overline{\D}^{c}\right>$.  We then
show that as soon as $\left<\tilde{\nu}^{c}\right> \neq 0$
inequality~(\ref{eq:parbkg})
emerges for a range of parameters in the theory.
\section{A problem in the Higgs sector: No parity violation}
\label{sec:min}
\begin{theorem}
\label{thm:triv}
The lowest state of the tree level SUSY Left - Right theory corresponds to
$\left < \D \right > = \left < \overline{\D} \right >
= \left < \D^{c} \right > = \left < \overline{\D}^{c} \right >
= \left < \Phi \right > = 0$
if $\left < \tilde{L}^{c} \right > = \left < \tilde{L} \right > = 0$,
except on one hypersurface in the parameter space which has a volume of measure
zero.
\end{theorem}
\pfhead
Let us consider equations~(\ref{eq:hig}) - (\ref{eq:dterm}) with $\tilde{L} =
\tilde{L}^{c} = 0$
Since we have $SU(2)_{L} \times SU(2)_{R}$ invariance we can diagonalize
$\Phi$ by using a bi-unitary transformation.  So we will work in the
$\Phi$ diagonal basis and let
\begin{equation}
\label{eq:phi}
\left<\Phi\right> = \left (\begin{array}{cc}
               \k & 0\\
                0 & \k'
             \end{array}
              \right )
\end{equation}
Further without loss of generality we can assume that $M'^2$ and $\m^2$
in equations~(\ref{eq:hig})-(\ref{eq:dterm})
are real and positive.  This is because we can reabsorb their phases into a
redefinition of the fields $\overline{\D}, \overline{\D}^{c}$ and $\Phi$
and other coupling constants.  $M_{1,2}^2$ and $M_{\phi}^2$
are real since the lagrangian
is real.  The key point now is to
recognize from equations~(\ref{eq:hig})-(\ref{eq:dterm})
that
for the potential to be bounded from below for the field going to infinity
the mass parameters must satisfy the constraints
\begin{equation}
\label{eq:delconst}
M^{2}_{1,2} \ge 0,
 ~~~~~~~~~~~~
M'^{2} \le M_{1} M_{2}
\end{equation}
\begin{equation}
\label{eq:phiconst}
M_{\phi}^{2} \ge 0
 ~~~~~and~~~~~~
\mu^{2} \le M_{\phi}^{2}.
\end{equation}
The first inequality in~(\ref{eq:delconst})
follows by looking at the directions $\left <\Phi\right > = 0$ and
$\left <\D\right > = \left <\D^{c}\right >
= v \tau_{1}$ and $\left <\overline{\D}\right >
= \left <\overline{\D}^{c}\right > = 0$
and vice versa.
The D-terms vanish in this direction and
unless the first inequality is satisfied
the potential is unbounded from below for $v \rightarrow \infty$.
(Since only $M_{1,2}^{2}$ appear in the potential we will hereafter
assume that $M_{1,2} \ge 0$ and real without loss of generality).  The
second inequality in ~(\ref{eq:delconst}) follows by looking along directions
$\left <\D^{c}\right > = \left <\D\right >
= {v \over M_{1}} \tau_{1}, ~\left <\overline{\D}\right >
= \left <\overline{\D}^{c}\right > =
- {v\over{M_{2}}} \tau_{1}$ and $\left <\Phi\right > = 0$.
Once again the D-terms vanish and the potential is unbounded unless
the inequality is satisfied.  The inequalities~(\ref{eq:phiconst})
follow similarly by looking along directions $\k' = - \k$ and real and the
rest of the fields equal to zero.
Inequalities~(\ref{eq:delconst}) and~(\ref{eq:phiconst})
imply that we can define angles
$\theta$ and $\theta'$ such that
\begin{equation}
\label{eq:angle}
M'^{2} = M_{1} M_{2} \cos{2\theta}
 ~~~~~~and~~~~~~~
\mu^{2} = M_{\phi}^{2} \cos2\theta'
\end{equation}
Since the vacuum expectation values
(VEVs) of all other fields other than the triplets and bidoublet are
zero the Higgs potential can be rewritten in terms of $\theta$ and $\theta'$
as
\begin{eqnarray}
V = & ~&cos^{2}\theta ~~Tr
\left ( M_{1} \D^{c} + M_{2} \overline{\D}^{c\dagger}
\right )^{\dagger}\left ( M_{1} \D^{c} + M_{2} \overline{\D}^{c\dagger}
\right ) \nonumber \\
& ~& + sin^{2}\theta ~~Tr
\left ( M_{1} \D^{c} - M_{2} \overline{\D}^{c\dagger}
\right )^{\dagger}\left ( M_{1} \D^{c} - M_{2} \overline{\D}^{c\dagger}\right)
\nonumber \\
& ~& + cos^{2}\theta'
M_{\phi}^{2}\left(\k + \k'^{*}\right)^{*} \left( \k + \k'^{*}\right)
+ sin^{2}\theta'
M_{\phi}^{2}\left(\k - \k'^{*}\right)^{*} \left( \k - \k'^{*}\right)
\nonumber \\
& ~& + \D \rightarrow \D^{c}, ~\overline{\D} \rightarrow \overline{\D}^{c}
\end{eqnarray}
Note that every term above is a norm and is hence positive semi-definite.
Thus it follows that the absolute minimum of the Higgs potential $V$ is
$V=0$ and this is obtained when
$\left < \D \right > = \left < \overline{\D} \right >
= \left < \D^{c} \right > = \left < \overline{\D}^{c} \right >
= \k = \k' = 0$.
\qed

A few comments are in order regarding the uniqueness of the absolute minimum.
The only case in which the fields can pick up VEVs and still have
$V = 0$, is if the following equations are satisfied:
\begin{equation}
\label{eq:hyper}
M'^{2} = M_{1} M_{2}
 ~~~~~and~~~~~
\mu^{2} = M_{\phi}^{2}.
\end{equation}
In this case there is a solution of
the form $\left <\D^{c}\right > = \left <\D\right > =
{v \over M_{1}} \tau_{1}, \left <\overline{\D}\right >
= \left <\overline{\D}^{c}\right > =
- {v\over{M_{2}}} \tau_{1}$ and $\k' = - \k$.
Such a solution is unstable under radiative corrections and in any case is
electric charge $(Q_{em})$ violating and parity conserving.
If we further restrict the coupling constant space such that $M_{1} = M_{2}$
as well as equations~(\ref{eq:hyper}) are satisfied, then there will be a
solution of the form $\D^{c} = - \overline{\D}^{c\dagger},
\D = - \overline{\D}^{\dagger}$ which
can be parity breaking and $Q_{em}$ conserving.  However these equality
relations for the mass-parameters correspond to a very constrained hypersurface
in the parameter space and will occur for points of measure zero in this space.
Further since they are not guaranteed by any symmetry of the theory,
such equality relations are unstable under radiative corrections.
Thus, the true minimum of the theory conserves parity.

In this section we have shown that the minimal SUSYLR model cannot
violate parity if it doesn't violate R-parity.  In the next section
we show that this conclusion holds at the tree level
for several extensions of the
minimal SUSYLR model as well.

\section{Non-Minimal SUSYLR models, the problem persists}
\label{sec:nonmin}
Certain extensions of minimal SUSLR models have been considered in
the literature.  A popular model is the SUSYLR model with two
Higgs bidoublet fields instead of one~$\left [ 11 \right ]$.
Another extension $\left [ 12 \right ]$ that has
been considered is the minimal SUSYLR model with an additional
parity odd singlet~$\left [ 13 \right ]$.
In this section we show that even for these models
the absolute minimum of the tree level Higgs potential cannot break
parity and conserve $Q_{em}$ if it does not break R-parity.

\subsection{Minimal SUSYLR + 1 extra bidoublet field}
\label{subsec:twobi}
Let
\begin{eqnarray}
\label{eq:twovac}
\left<\Phi_{1}\right> = & ~& \left ( \begin{array}{cc}
                  \k_{1} & 0 \\
                   0    &  \k'_{1}
                   \end{array}
                \right )
 ~~~~~\left<\Phi_{2}\right> = \left ( \begin{array}{cc}
                  \k_{2} & 0 \\
                   0    &  \k'_{2}
                   \end{array}
                \right )
\end{eqnarray}
be the VEVs of  the bidoublet fields.  We choose this form since it is
the most general form consistent with conserving $Q_{em}$.
The Higgs potential of the triplet and the bidoublet
fields\footnote{note that the
VEVs of the slepton fields are set equal to zero so that
R-parity remains unbroken} is of the form:
\begin{eqnarray}
\label{eq:twobi}
V_{Higgs} =& ~& V\left(\D, \overline{\D},
\k_{1}, \k'_{1}, \k_{2}, \k'_{2}\right)
 ~+ ~V\left(\D^{c}, \overline{\D}^{c},
\k_{1}, \k'_{1}, \k_{2}, \k'_{2}\right) \nonumber \\
& ~& + {{g'^{2}} \over 2} \left(Tr\left(\D^{\dagger}\D - \D^{c\dagger}\D^{c}
- \overline{\D}^{\dagger}\overline{\D}
+ \overline{\D}^{c\dagger}\overline{\D}^{c}\right)\right)^{2}
\end{eqnarray}
Note that the above potential is invariant under $\D \leftrightarrow \D^{c},
 ~\overline{\D} \leftrightarrow \overline{\D}^{c}.$ Further if we set
$g' = 0$, then for {\it any} value of $\k_{1},\k'_{1}, \k_{2}, \k'_{2}$,
the value of the triplet fields that minimize the potential
of equation~(\ref{eq:twobi})
absolutely is
$\D = \D^{c}, ~\overline{\D} = \overline{\D}^{c}$ due
to symmetry.  Now let $g'^{2} > 0$ and arbitrary.  The last term in
the Higgs potential of equation~(\ref{eq:twobi}) is positive semi-definite
and for $\D = \D^{c}, ~\overline{\D} = \overline{\D}^{c}$
it also has its lowest value which is zero.
{}From this it follows that at the absolute
minimum $\D = \D^{c}, ~\overline{\D} = \overline{\D}^{c}$ and parity remains
unbroken!  Note that the same argument can be made no matter how many
triplet or bidoublet fields we have in the SUSYLR theory as long as they
are the only fields that pick up VEVs.
We also note that as in section~\ref{sec:min} there is a hypersurface
in the parameter space, where parity can break but this is unstable under
radiative corrections.
\subsection{Minimal SUSYLR + Parity Odd Singlet}
\label{subsec:odd}
Let $\s$ be the Parity odd singlet\footnote{under Parity $\s \rightarrow - \s$
.}.  Now the superpotential $W$
has the following additional terms~$\left [ 12 \right ]$:
\begin{equation}
W = A \s Tr \left( \D \overline{\D} - \D^{c}\overline{\D}^{c}\right) + W(\s)
\end{equation}
The only additional term that this introduces in the Higgs potential
that is not of the form given below in equation~(\ref{eq:comb}) is
\begin{equation}
\label{eq:Fterm}
V_{Higgs}~the ~new~F-~term = ~Tr \mid f(\s)
+ A \left( \D \overline{\D} - \D^{c}\overline{\D}^{c}\right)\mid^{2}
\end{equation}
where $f(\s)$ is a function of $\s$ alone.
The key is to note that $\s$ does not couple directly to $\Phi$ and
therefore inequality~(\ref{eq:phiconst}) still holds.  This implies that
the sum of the quadratic terms
involving  $\Phi$ alone are positive semi definite.  We also
note that the D-terms of the Higgs potential are all
positive semi-definite.  Thus the only way the Higgs potential can
attain a negative value at the minimum is if some of the the quadratic and
the cubic terms involving
the triplet and/or  singlet fields are negative at the absolute minimum.
The {\it only} way in which the triplet fields occur in these terms
is in the combination
\begin{equation}
\label{eq:comb}
Tr \left(\D^{\dagger} \D \right ),
Tr \left(\overline{\D}^{\dagger} \overline{\D} \right ),
Tr \left(\D^{c\dagger} \D^{c} \right ),
Tr \left(\overline{\D}^{c\dagger} \overline{\D}^{c} \right ),
Tr \left(\D \overline{\D}\right)~~and ~~Tr \left(\D^{c}
\overline{\D}^{c}\right)
\end{equation}
and their products with the singlet field.
We will now show that given any configuration of the triplet fields that
conserves $Q_{em}$ there is a configuration that violates $Q_{em}$
for which the value of these terms is {\it unchanged}.
The most general form of the triplet fields that conserves
$Q_{em}$ is
\begin{eqnarray}
\label{eq:delform}
\left<\D^{c}\right> = & &\left (\begin{array}{cc}
                    0 & 0 \\
                    \d' & 0
                   \end{array}
                    \right )
 ~~~ \left<\D\right> =  \left (\begin{array}{cc}
                    0 & 0 \\
                    \d & 0
                   \end{array}
                    \right )\nonumber\\
\left<\overline{\D}^{c}\right> = & &\left (\begin{array}{cc}
                    0 & \overline{\d}' \\
                    0 & 0
                   \end{array}
                    \right )
 ~~~ \left<\overline{\D}\right> =  \left (\begin{array}{cc}
                    0 & \overline{\d} \\
                     0 & 0
                   \end{array}
                    \right ).
\end{eqnarray}
It is easy to check that the following $Q_{em}$ violating form preserves
the values of the terms in equation~(\ref{eq:comb}):
\begin{eqnarray}
\label{eq:viol}
\left<\D^{c}\right> = & &{1 \over { \sqrt{2} }} \left (\begin{array}{cc}
                    0 & \d' \\
                    \d' & 0
                   \end{array}
                    \right )
 ~~~ \left<\D\right> =  {1 \over { \sqrt{2} }}\left (\begin{array}{cc}
                    0 & \d \\
                    \d & 0
                   \end{array}
                    \right )\nonumber\\
\left<\overline{\D}^{c}\right>
= & &{1 \over { \sqrt{2} }}\left (\begin{array}{cc}
                    0 & \overline{\d'} \\
                    \overline{\d}' & 0
                   \end{array}
                    \right )
 ~~~ \left<\overline{\D}\right>
= {1 \over { \sqrt{2} }} \left (\begin{array}{cc}
                    0 & \overline{\d} \\
                   \overline{\d} & 0
                   \end{array}
                    \right ).
\end{eqnarray}
In addition by substituting the above
$Q_{em}$ violating form in the D-terms of the Higgs potential it is
easy to see that together with the choice $\Phi = 0$,
all the positive semi-definite D-terms and quadratic terms in $\Phi$
that we referred to earlier take their lowest possible value which is
zero.  The only other term that remains is the
F-term in equation~(\ref{eq:Fterm})
and it is easy to see that this term is minimized too!!
Thus the $Q_{em}$ violating configuration has a lower value for the
potential than the $Q_{em}$ conserving form\footnote{except on a hypersurface
in the parameter space of the Higgs potential where the two forms
have the same values for the potential.  The
reader is referred to the discussion at the end of
section~\ref{sec:min} for such a case.}. Thus
we conclude that the absolute minimum  of the tree level Higgs potential
violates $Q_{em}$ if R-parity is not broken.

In the appendix we investigate some more extensions of the minimal SUSYLR
model and show that minimal SUSYLR + Extra bidoublets + Parity odd singlet
also has the same problem of violating $Q_{em}$.  We further show in the
appendix
that minimal SUSYLR + Parity Even Singlet cannot break parity without
breaking R-parity either. In the next section we show that if we
violate R-parity then in the minimal SUSYLR model itself, we can obtain
absolute minima that violate parity and preserve $Q_{em}$.

\section{R - Parity violation cures the problem}
\label{sec:rpar}
In this section we show that
by giving $\tilde{\n}^{c}$ a VEV and thereby breaking
R-parity spontaneously, we can salvage the minimal SUSYLR theory.
We will further demonstrate that
we can have a stable non-trivial solution\footnote{by this we mean that
at least some fields pick up non-zero VEVs and there are no directions in the
tree-level potential wherein $V \rightarrow - \infty$} for the absolute
minimum which conserves $Q_{em}$ at the tree level itself without needing
quantum corrections for stability.  This is a nice feature since it is present
not
only  in all non-SUSY models but also in SUSY $SU(2)_{L} \times
U(1)_{Y}$ as well.

We will now find a region in the coupling constant space such that the
absolute minimum for the tree level Higgs potential breaks parity but
conserves $Q_{em}$.  We will do this in perturbation by treating certain
parameters as small.

Let us consider
the Higgs potential in equations~(\ref{eq:hig}) - (\ref{eq:dterm})
and let us choose a region
in coupling constant space such that $g^{2}$ and $g'^{2}$ are much smaller
than the other dimensionless coupling constants $h^{2}$ and $f^{2}$.
This ensures that the troublesome D-terms that were responsible for
breaking $Q_{em}$ and preserving parity are weaker than the quartic terms and
tri-linear terms between the triplet and slepton fields.  Also,
in order to
achieve
the hierarchy $\left<L^{c}\right>,
\left<\D^{c}\right> ~ >> ~ \left<\Phi\right>$
we will require that the coupling constants and mass terms involving
$\Phi$ are smaller than those not involving them.  Under these
assumptions the
potential in equations~(\ref{eq:hig}) - (\ref{eq:dterm}) simplifies to:
\begin{eqnarray}
\label{eq:snupot}
V = & ~& m_{l}^{2} \left (
\tilde{L}^{\dagger} \tilde{L} + \tilde{L}^{c\dagger} \tilde{L}^{c} \right )
+ M_{1}^{2} ~Tr \left ( \D^{\dagger} \D + \D^{c\dagger} \D^{c} \right )
+ M_{2}^{2} ~Tr \left ( \overline{\D}^{\dagger} \overline{\D}
+ \overline{\D}^{c\dagger} \overline{\D}^{c} \right ) \nonumber \\
& ~& + \mid h \mid^{2} \tilde{L}^{c\dagger}
\tilde{L}^{c} \tilde{L}^{\dagger} \tilde{L} + \mid f \mid^{2}
\left ( \left ( \tilde{L}^{\dagger} \tilde{L} \right )^{2} +
\left (\tilde{L}^{c\dagger} \tilde{L}^{c}\right )
^{2} \right )\nonumber \\
& ~& + 4 \mid f \mid^{2} \left ( \mid \tilde{L}^{cT} \tau_{2} \D^{c} \mid^{2}
+ \mid \tilde{L}^{T} \tau_{2} \D \mid^{2}\right ) + M'^{2}~Tr\left (
\D \overline {\D} + \D^{c} \overline{\D}^{c} + h.c. \right )
\nonumber \\
& ~& + \left (\tilde{L}^{T} \tau_{2} \left (
i v \D + i M^{*} f \overline{\D}^{\dagger} \right ) \tilde{L}
+ \tilde{L}^{cT} \tau_{2} \left ( i v \D^{c} +
i M^{*} f \overline{\D}^{c\dagger} \right ) \tilde{L}^{c} + h.c. \right )
\end{eqnarray}
In the above we will require
that inequalities~(\ref{eq:delconst}),
(\ref{eq:phiconst}) and $\mid M \mid < M_{2}$
be satisfied
so that there is no direction where $V \rightarrow - \infty$.
We will now use the $SU(2)_{L} \times SU(2)_{R}$ invariance to
choose
\begin{equation}
\label{eq:snuform}
\left<\tilde{L}^{c}\right> = \left ( \begin{array}{c}
             l' \\
             0
         \end{array} \right )
 ~~~,~~~\left<\tilde{L}\right> = \left ( \begin{array}{c}
             l \\
             0
         \end{array} \right )
\end{equation}
with $l$ and $l'$ real and positive.
Now let the triplet fields have their most general form
\begin{equation}
\left<\overline{\D}^{c}\right> = \left ( \begin{array}{cc}
                               {a \over {\surd 2}} & \overline{\d}'\\
                                    b    &  {{- a} \over {\surd 2}}
                                    \end{array}
                                    \right )
 ~~~~and ~~~~\left<\D^{c}\right> = \left ( \begin{array}{cc}
                               {c \over {\surd 2}} & d\\
                                    \d'    &  {{- c} \over {\surd 2}}
                                    \end{array}
                                    \right )
\end{equation}
Substituting these VEVs into equation~(\ref{eq:snupot}) it is easy to see that
the only terms involving $a, b, c,$ or $d$ are
\begin{eqnarray}
& ~&M_{1}^{2} \left ( \mid c\mid^{2} + \mid d \mid^{2} \right )
+ M_{2}^{2} \left ( \mid a\mid^{2} + \mid b \mid^{2} \right ) \nonumber \\
& ~& + \left ( M'^{2} \left ( ac + bd \right ) + h.c. \right )
+ 2 \mid f \mid^{2} \mid c \mid^{2} l'^{2}
\end{eqnarray}
which are minimized when $a = b = c = d = 0$
due to inequalities~(\ref{eq:delconst}).
By similar arguments we can show that
for the absolute minimum, all the
triplet fields {\it must} have the $Q_{em}$ conserving
form of equation~(\ref{eq:delform})
The task now is to determine the unknown VEVs in equations~(\ref{eq:snuform})
and~(\ref{eq:delform})
by substituting them into equation~(\ref{eq:snupot})
and minimizing the potential. Since the equations for the minima will
be coupled cubic equations, they are a bit complicated in general.
Therefore we will solve the equations by treating
$M_{1}^{2}$ and $M'^2$
as perturbations that are smaller than the other mass parameters and
will neglect them in what follows. Also we will assume that all coupling
constants, including $v$ and $f$ are real.
These requirements, though not essential, will
considerably simplify the mathematics and keep it from being messy.
Also note that our main goal is to prove that there exists at least  {\it a}
region in the parameter space where
the absolute minimum breaks parity and conserves $Q_{em}$.
Thus the
potential in equation~(\ref{eq:snupot}) becomes
\begin{eqnarray}
\label{eq:simppot}
V = & ~& m_{l}^{2} \left (l'^{2} + l^{2} \right ) +
M_{2}^{2} \left ( \mid \overline{\d}'^{2}\mid
+ \mid \overline{\d}^{2} \mid \right )
+ f^{2} \left ( l'^{4} + l^{4} \right ) \nonumber \\
& ~& + 4 f^{2} \left ( l'^{2} \mid \d'^{2}\mid
+ l^{2} \mid \d^{2}\mid \right ) + h^{2} l'^{2} l^{2} \nonumber \\
& ~& + \left ( v \left (l'^{2} \d' + l^{2} \d \right ) + f M \left (
l'^{2} \overline{\d}' + l^{2}
\overline{\d} \right ) + complex ~conjugate \right ).
\end{eqnarray}
The first thing to note is that since
the only terms in equation~(\ref{eq:simppot})
that care about the phases of the fields are the trilinear terms,
all fields will pick up real VEVs.  This is because
if $v$ is positive (note that it is real by assumption), $\d'$ will
minimize the potential by being real and negative (rather than having
a complex phase) and if $v$ is negative, $\d'$ will be positive.
The same argument goes through for other fields and hence in what follows
we will keep all fields real.
The conditions for the extrema of equation~(\ref{eq:simppot}) are:
\begin{eqnarray}
\label{eq:extrem}
{{\partial V} \over {\partial l'}} = & ~& \left ( m_{l}^{2} +
4 f^{2} \d'^{2}
+ h ^{2} l^{2} + 2 v \d'
+ 2 f M \overline{\d}' \right ) l' + 2 f^{2} l'^{3} = 0 \nonumber \\
{{\partial V} \over {\partial \d'}} = & ~& 4 f^{2} l'^{2} \d' + v l'^{2} = 0
\nonumber \\
{{\partial V} \over {\partial \overline{\d}'}} = & ~&
M_{2}^{2} \overline{\d}' + f M l'^{2} = 0
\end{eqnarray}
and three corresponding equations with $l' \leftrightarrow l,
 ~\d' \rightarrow \d$ and $\overline{\d}' \rightarrow \overline{\d}.$
The solutions for the extrema are:
\begin{eqnarray}
\label{eq:trivsol}
The ~ trivial ~ solution\footnotemark: & ~& l = l' =
\overline{\d}' = \overline{\d} = 0
 ~~~ for~ which ~ V = V_{TS} = 0
\end{eqnarray}
\begin{eqnarray}
\label{eq:pbs}
The ~parity ~breaking ~solution\footnotemark: & ~& l = \overline{\d} =  0, ~
\d' = - {v \over {4 f^{2}}}, ~
\overline{\d}' = - {{f M l'^{2}} \over {M_{2}^{2}}}, \nonumber \\
& ~& l'^{2} = l_{PBS}'^{2} =
{{\left(v^{2} - 4 f^{2} m_{l}^{2} \right ) M_{2}^{2}} \over {8 f^{4} \left (
M_{2}^{2} - M^{2} \right )}} \nonumber \\
& ~& V = V_{PBS}= - f^{2} \left ( 1 - {{M^{2}} \over
{M_{2}^{2}}} \right )l_{PBS}'^{4}
\end{eqnarray}
\begin{eqnarray}
\label{eq:pps}
The ~parity ~preserving ~solution: & ~& \d' = \d = - {v \over {4 f^{2}}}, ~
\overline{\d}' = \overline{\d}
= - {{f M l'^{2}} \over {M_{2}^{2}}}, \nonumber \\
& ~& l'^{2} = l^{2} = l_{PPS}'^{2}
=  {{\left(v^{2} - 4 f^{2} m_{l}^{2} \right ) M_{2}^{2}} \over {8 f^{4} \left (
M_{2}^{2} \left( 1 + {h^{2} \over {2 f^{2}}} \right )
- M^{2}  \right )}} \nonumber \\
& ~& V = V_{PPS}= - 2 \left (f^{2}
\left ( 1 - {{M^{2}} \over {M_{2}^{2}}} \right ) + {{h^{2}} \over 2} \right )
l_{PPS}'^{4}.
\end{eqnarray}
\addtocounter{footnote}{-1}\footnotetext{the
alert reader will notice that when the perturbation
$M_{1}^{2} > 0$ is turned on it will force $\d = \d' = 0$.}
\addtocounter{footnote}{1}\footnotetext{the perturbation $M_{1}^{2} > 0$ will
imply that $\d = 0$ for the minimum}
Note that in the above the parity breaking solution will exist if
\begin{equation}
\label{eq:triconst}
v^{2} - 4 f^{2} m_{l}^{2} > 0.
\end{equation}
Further the parity breaking solution will be the absolute minimum if
\begin{equation}
\label{eq:absconst}
V_{PBS} ~<~ V_{PPS}.
\end{equation}
Substituting for $V_{PBS}$ and
$V_{PPS}$ from equations~(\ref{eq:pbs}) and~(\ref{eq:pps})
we find that inequality~(\ref{eq:absconst}) is satisfied if
\begin{equation}
h^{2} ~> ~2 f^{2} {{\left( M_{2}^{2} - M^{2} \right )} \over {M_{2}^{2}}}
\end{equation}
Thus we have obtained a region in parameter space such that parity is
broken.

If we substitute the VEVs given by equations~(\ref{eq:pbs}) into the
Higgs potential in equations~(\ref{eq:hig}) - (\ref{eq:dterm}),
then they will act as sources for
the $\Phi$ - field
to pick up a VEV.  We will now show that $\left<\Phi\right>$ at
the absolute minimum doesn't break $Q_{em}$.

Since $g^{2}, g'^{2} << h^{2}$ the most dominant coupling term between the
$\Phi$ and the other right-handed fields is the positive definite term
\begin{equation}
\label{eq:domphi}
h^{2}
\left<\tilde{L}^{c^{\dagger}}\right> \tau_{2} \Phi^{\dagger} \Phi \tau_{2}
\left<\tilde{L}^{c}\right>
\end{equation}
Let $\left<\Phi\right>$ take its most general form given by:
\begin{equation}
\label{eq:phigen}
\left<\Phi\right> = \left ( \begin{array}{cc}
                               \k & a \nonumber\\
                                b & \k'
                                 \end{array}
                              \right )
\end{equation}
Substituting for $\tilde{L}^{c}$
from equations~(\ref{eq:snuform}) and~(\ref{eq:pbs}) into~(\ref{eq:domphi})
and minimizing we obtain $\k' = a = 0$.  The only other term
in equation~(\ref{eq:hig}) - (\ref{eq:dterm}) that
couples $\Phi$ to the VEV of the other right handed fields is the
cross term in the D-term of the Higgs potential which on substitution is:
\begin{equation}
\label{eq:phisource}
{{g^{2}} \over 4}
\left( l'^{2} - 2 \d'^{2} + 2 \overline{\d}'^{2} \right ) \left ( \mid
\k \mid^{2} + \mid b \mid^{2} \right )
\end{equation}
If\footnote{this condition can be easily arranged by choosing $m_l^{2}$
positive and large enough as can be seen from equation~(\ref{eq:pbs}).}
$2 \d'^{2} >  l'^{2} + 2 \overline{\d}'^{2}$ this acts as a source
for $\mid \k \mid^{2} + \mid b \mid^{2}$ to pick up a VEV.  At this
stage there are two degenerate absolute minima of the form $\k \neq 0~
and ~b = 0$ or $b \neq 0 ~and~ \k = 0$.  Either $\k$ or $b$ is zero
because if both have
non-zero values the D-terms will contribute extra positive amounts
to the potential due to cross terms between them. The reason the
absolute minimum has $\k \neq 0$ and $b = 0$ is because the tri-linear terms
between $\Phi, \tilde{L}$ and $\tilde{L}^{c}$ couple only to the
electrically neutral component $\k$ and not $b$.  The phase of $\k$
will be determined by these terms
so that they contribute the maximal negative amount
to the potential and this will split the
degeneracy between the two minima in favour of the $Q_{em}$ conserving one.
Note that a small value of $\tilde{L}$ itself is induced in this process.
$\k$ can be easily
calculated by writing down all dominant terms in the Higgs potential
as follows:
\begin{equation}
\label{eq:k}
V\left (terms ~involving ~ \k \right ) = {{g^{2}} \over 4} \left (
\left( 4 {{M_{\phi}^{2}} \over {g^{2}}} +
l'^{2} - 2 \d'^{2} + 2 \overline{\d}'^{2} \right )  \mid \k \mid^{2}
+ \mid \k \mid^{4}\right )
\end{equation}
\vskip0.5in
Minimizing with respect to $\k$ we obtain
\begin{equation}
\label{eq:k'value}
\mid \k \mid^{2} = \d'^{2} - \overline{\d}'^{2} - {1 \over 2} l'^{2}
- 2 {{M_{\phi}^{2}} \over {g^{2}}}
\end{equation}
Note that inorder that $\k << \d'$ we have the usual problem of fine-tuning
the the terms in equation~(\ref{eq:k}).
Once $\k$ picks up a VEV it acts as a source for $\k'$ (and not $a$ or $b$)to
pick up a VEV due to the term ${{\m^{2}} \over 2} Tr \tau_{2} \Phi^{T}
\tau_{2} \Phi$ and we can find the values of $\k'$ by minimizing the
dominant terms namely,
\begin{equation}
\label{eq:keqn}
V \left(dominant ~terms ~involving ~\k' \right ) =
\mu^{2} \k \k' + h.c. + \left (M_{\phi}^{2} + h^{2} l'^{2} \right ) \mid
\k'\mid^{2}
\end{equation}
A point worth noting is
the hierarchy $\k' << \k$ is automatically obtained in this mechanism.

Finally, we note that $l, \d, \overline{\d}$ will
pick up small VEVs due to the trilinear couplings with the fields
that have already picked up VEVs.
The reason these fields will pick up $Q_{em}$ conserving VEVs is that
only the electrically neutral components of these fields
couple to the other VEVs and so only they can be induced.

Thus we have shown that there exists a region in parameter space of
the Higgs potential for minimal SUSLR model such that parity and R-parity
are broken
and $Q_{em}$ is conserved. In the process we have obtained several
inequality relations between the coupling constants in the Higgs
potential and they can be useful for phenomenology.
\section{Effect of Radiative Corrections}
\label{sec:rad}
In section 3 we showed that at tree level there is no parity
violation without R-parity violation
in the minimal SUSYLR model.  In this section we will
argue that this result is true even after radiative corrections are taken
into account. Since for phenomenological reasons
$\left <\Phi\right > << \left <\D^{c}\right >$,
we will set $\left < \Phi \right > = 0$
and will only consider VEVs for the triplets\footnote{We
have checked that an argument similar to the one in this section
can be made even if $\left<\Phi\right> \neq 0$}.
The effective potential after radiative corrections are taken into account
is:
\begin{equation}
\label{eq:veff}
V_{eff} = V + V_{rad}\left(Q\right)
\end{equation}
where $V$ is the tree level higgs potential in equation~(\ref{eq:hig})
and $V_{rad}\left ( Q \right )$ is the radiative correction evaluated
using the usual Coleman-Weinberg technique~$\left [ 14 \right ]$
at a renormalization
scale $Q$. Note that all parameters of the theory
depend on $Q$ such that
$V_{eff}$ is independent of $Q~ \left [ 15, 16 \right ]$.  In other words,
\begin{equation}
\label{eq:ren}
{{d V_{eff}} \over {d Q}} = 0
\end{equation}
where the mass parameters and the coupling constants of the theory are
functions of $Q$.
Equation~(\ref{eq:ren}) implies that we can choose any convenient value
of $Q$ to evaluate $V_{eff}$.
One of the standard ansatz~$\left [ 15, 16 \right ]$
in supersymmetric theories is to choose
$Q$ such that
\begin{equation}
\label{eq:pres}
{\left .{{\partial V_{rad}\left ( Q \right )} \over {\partial \chi}}
\right|}_{\chi = \left < \chi \right >, rest~of~fields~at~their~vevs~too} = 0
\end{equation}
where $\chi$ is a scalar field (or a linear combination of scalar fields)
in the theory. Note that
the mass parameters
and other coupling constants of the theory are now functions of $\hat{Q}$ where
$Q = \hat{Q}$ solves equation~(\ref{eq:pres}).
Physically equation~(\ref{eq:pres}) implies that if we choose a configuration
infinitesimally
different from the vacuum configuration for the field $\chi$, then
$Q$ is chosen such that the radiative corrections for the two configurations
are the same.  For our problem we will use a prescription slightly different
from equation~(\ref{eq:pres}) but which is physically very similar.

Let us assume that the triplet fields pick up $Q_{em}$ conserving
VEVs of the form given by~(\ref{eq:delform})
even after radiative corrections are taken into account.
We will call this $Q_{em}$ conserving configuration
of fields as Configuration $1$. Let us consider
a slightly different
$Q_{em}$ violating configuration (Configuration 2) namely,
\begin{equation}
\label{eq:diffdel}
\D^{c} = \d' \left ( \begin{array}{cc}
                      0 & sin \theta \\
                     cos \theta & 0
                           \end{array}
                       \right ),
{}~~~\overline{\D}^{c} = \overline{\d}'   \left ( \begin{array}{cc}
                      0 & cos \theta \\
                     sin \theta & 0
                           \end{array}
                       \right )
\end{equation}
where $\theta$ is any {\it fixed} angle.
Let us choose $Q$ such that the radiative corrections
($V_{rad}$ in equation~(\ref{eq:veff})) for Configuration 1 and Configuration 2
are equal at $Q = \hat{Q}$.
This is our renormalization prescription. Of course as discussed in
the previous paragraph all coupling constants in the tree level
Higgs potential in equations~(\ref{eq:hig}) - (\ref{eq:dterm}) are now
functions of $\hat{Q}$.  Let us now compare the values of the effective
potential $V_{eff}$ in equation~(\ref{eq:veff}) for the two configurations.
Our renormalization prescription immediately implies that
\begin{equation}
V_{eff} \left ( Configuration ~2 \right ) - V_{eff} \left ( Configuration ~1
\right ) = V \left (Config. 2 \right ) - V \left ( Config. 1 \right )
\end{equation}
Substituting the forms of the triplet fields for both configurations
into $V$ given by equations~(\ref{eq:hig}) - (\ref{eq:dterm}), it is easy to
see that above difference is negative no matter what
values the coupling constants
and mass parameters in
the Higgs potential $V$ take.
Note that we don't even have to assume that
inequalities~(\ref{eq:delconst}) and~(\ref{eq:phiconst})
are satisfied. However we require $g^{2}\left ( \hat{Q} \right ) > 0$ which
is actually a requirement from gauge invariance since $g$ is a gauge
coupling constant and hence is purely real.
Once again it is the
D - term whose value is lower for the $Q_{em}$ violating configuration
while all other terms have the same value for both Configuration 1 and 2.

Thus choosing a suitable renormalization prescription
we have argued that in the minimal SUSLR model
$Q_{em}$ is violated if R- parity is not broken even
after quantum effects are taken into account.
\section{Conclusion}
In summary, we have shown that in a class of minimal supersymmetric left
right models where R-parity symmetry is automatic in the symmetry limit,
spontaneous breakdown of parity {\it requires} spontaneous breakdown
of R-parity.
We have also argued that this result established for the tree level
potential is unlikely to be effected when one-loop effects are included.
This intriguing
result connects two physically
different scales in supersymmetric models.  Its phenomenological
implications will be the subject of a future
publication.
\vskip0.2in
\begin{center}
{\Large\bf Acknowledgement}
\end{center}
We would like to thank K. S. Babu for discussions. One of us (RK) would
also like to thank Sanjay Kodiyalam and Lubna Rana of
the University of Maryland
and Ramesh Kumar Sitaraman of the Princeton University
for insightful discussions.
This work was supported by NSF Grant No. PHY - 9119745.

\appendix
\section{Appendix}
In this appendix we prove that minimal SUSYLR + Extra bidoublets +
Parity odd singlet also has the unsatisfactory feature of having only
$Q_{em}$ violating absolute minimum if R-parity is not broken.
We also show that
for minimal SUSYLR + Parity even singlet model the
value of the tree-level potential for parity conserving
VEVs is always lower than the value for VEVs that violate Parity maximally
if R-parity is unbroken.

\subsection{Minimal SUSYLR + Extra bidoublets + Parity odd singlets}
The terms in the Higgs potential that involve
the bidoublets
are of the form $Tr \left (\Phi_{i}^{\dagger}\Phi_{j} \right )$,
$Tr \left ( \tau_{2}
\Phi_{i}^{T} \tau_{2} \Phi_{j}\right )$ where the indices $i$ and $j$
run over the bidoublets in the theory
and the D-terms.  To keep things simple
let us consider two bidoublets and let them take
$Q_{em}$ conserving VEVs of the form in equation~(\ref{eq:twovac}).
Note that the above mentioned terms are all invariant under
$\k_{1} \leftrightarrow \k'_{1}, \k_{2} \leftrightarrow \k'_{2}.$
Due to this symmetry it is easy to see that the most
general tree-level higgs potential
involving the bidoublets alone can be written in the following
matrix notation.
\begin{equation}
\label{eq:mat}
V =   \left ( \k ~ \k' \right )^{\dagger}
                 \left ( \begin{array}{cc}
                         M & B \\
                         B & M
                          \end{array} \right )
        \left ( \begin{array}{c}
                     \k \\
                      \k'
                  \end{array}
               \right )
 ~~+~~ {{g^{2}}
\over 4} \left ( \k^{\dagger} \k - \k'^{\dagger} \k'\right )^{2}
\end{equation}
where the $4 \times 4$ matrices $M$ and $B$ are hermitian and
\begin{equation}
\label{eq:K}
\k^{\dagger} = \left ( \k^{*}_{1} ~ \k^{*}_{2} ~ \k_{1} ~ \k_{2} \right ),
 ~~~ \k'^{\dagger}
= \left ( \k'^{*}_{1} ~ \k'^{*}_{2} ~ \k'_{1} ~ \k'_{2} \right ).
\end{equation}
In the
direction $\k' = \pm \k,$ the quartic term in equation~(\ref{eq:mat}) vanishes.
This implies that if the potential is to be bounded from going to $- \infty$
the mass matrix in equation~(\ref{eq:mat})
must satisfy the following constraints for any arbitrary $\k$:
\begin{eqnarray}
\label{eq:matconst}
   \left ( \k ~ \k \right )^{\dagger}
                 \left ( \begin{array}{cc}
                         M & B \\
                         B & M
                          \end{array} \right )
        \left ( \begin{array}{c}
                     \k \\
                      \k
                  \end{array}
               \right ) \nonumber & ~& \ge ~0\\
   \left ( \k ~ - \k \right )^{\dagger}
                 \left ( \begin{array}{cc}
                         M & B \\
                         B & M
                          \end{array} \right )
        \left ( \begin{array}{c}
                     \k \\
                      - \k
                  \end{array}
               \right ) & ~& \ge ~0.
\end{eqnarray}
Now note that any {\it general} configuration $\left ( \k ~ \k' \right )$
can be written as:
\begin{equation}
\label{eq:general}
     \left ( \begin{array}{c}
            \k \\
            \k'
         \end{array} \right )
 ~~= ~~ {1 \over 2} \left ( \begin{array}{c}
                          \k + \k' \\
                          \k + \k'
                          \end{array} \right )
  ~~+ ~~{1 \over 2} \left ( \begin{array}{c}
                          \k - \k' \\
                          - \k + \k'
                          \end{array} \right )
\end{equation}
Substituting the above in equation~(\ref{eq:mat}) it is easy to see that
the contribution from the quadratic terms in $\k$ and $\k'$ is always
positive semi- definite due to inequalities~(\ref{eq:matconst}).
This
was the only information from the $\Phi$ sector which was needed to prove
that the absolute minimum violates $Q_{em}$ in subsection~\ref{subsec:odd}.
Thus following the same steps as in subsection~\ref{subsec:odd} we can conclude
that no matter how many bidoublets or parity odd singlets we have,
$Q_{em}$ is violated if R - parity is not broken.

\subsection{Minimal SUSYLR + Parity Even Singlet}
Let $\o$ be the parity even singlet~\footnote{under Parity $\o \rightarrow
\o$.}. The superpotential $W$ has the following additional terms:
\begin{equation}
W = B \o Tr \left (\D\overline{\D} + \D^{c} \overline{\D}^{c}\right ) +
C \o Tr \left ( \tau_{2} \Phi^{T} \tau_{2} \Phi \right ) + W\left ( \o
\right ).
\end{equation}
The only additional term that this introduces in the Higgs potential
that is not quadratic in the triplet or bidoublet fields
is
\begin{equation}
\label{eq:Feven}
V_{new~F - ~term} = Tr \mid f \left ( \o \right ) +
B \left ( \D \overline{\D} + \D^{c} \overline{\D}^{c} \right)
+ C ~\tau_{2} \Phi^{T} \tau_{2} \Phi \mid^{2}
\end{equation}
Consider a configuration that violates parity maximally namely,
$\D = \overline{\D} = 0$ and $\D^{c} = X, \overline{\D}^{c} = Y$,
where X and Y are $Q_{em}$ conserving VEVs.  We will
show that regardless of the VEV of the $\Phi$ field there exists
a parity conserving configuration which lowers the value of the
tree level configuration\footnote{except on one hypersurface
in the parameter space where the two configurations have the same
value for the potential.}.  The reader may verify that this configuration
is $\D = \D^{c} = X / \surd 2$ and $\overline{\D} = \overline{\D}^{c} =
Y / \surd 2$.  Basically the parity conserving configuration lowers
the values of the positive semi-definite
D - terms, regardless of the value of the
diagonal $\Phi$ field.  The values of the quadratic terms
and the new F - term are
invariant for the two configurations.

We will end this sub-section here though the proof
only considered initial configurations that violated Parity
maximally and did not consider
cases where $\left<\D\right> \neq 0$ initially.
Since phenomenologically
$\left<\D\right> << \left<\D^{c}\right>$
our assumption that for parity
violating configuration $\D = 0$ is not too bad.

Our conclusion then is that
if we insist on a SUSYLR  theory with triplet fields to ensure
the see-saw mechanism for the neutrino masses and keep R-parity
intact, then
we should go well beyond the
minimal models
in order to achieve breakdown of parity.
\newpage
\begin{center}
{\Huge\bf References}
\end{center}
\begin{tabbing}
$\left [1 \right ] ~$ \= L. Hall and M. Suzuki,
{\it Nucl. Phys.} {\bf B 231}, 419 (1984)\\
   \> I. H. Lee, {\it Nucl. Phys.} {\bf B 246}, 120 (1984)\\
   \> F. Zwirner, {\it Phys. Lett.} {\bf 132B}, 103 (1983) \\
   \> S. Dawson, {\it Nucl. Phys.} {\bf B 261}, 297 (1985) \\
   \> R. N. Mohapatra, {\it Phys. Rev.} {\bf D 34}, 3457 (1986) \\
   \> V. Barger, G.F. Giudice and
T. Han, {\it Phys. Rev.} {\bf D 40}, 2987 (1989) \\
$\left [ 2 \right ]$ \> B. Campbell,
S. Davidson, J. Ellis and K. Olive, {\it Phys. Lett.} {\bf 256B}, 457 (1991)\\

$\left [ 3 \right ]$ \> R. N. Mohapatra, ref. 1 \\
   \> A. Font,
L.E. Ibanez and F. Quevedo, {\it Phys. Lett.} {\bf B 228} 79 (1989) \\
$\left [ 4 \right ]$ \> For a
general
study of this see S. P. Martin, {\it Phys. Rev.} {\bf D 46}, 2769 (1992)\\
$\left [ 5 \right ]$ \> C. S. Aulakh
and R. N. Mohapatra, {\it Phys. Lett.} {\bf 119B}, 136 (1983)\\
$\left [ 6 \right ]$ \> G. G. Ross and J. W. F. Valle,
{\it Phys. Lett.} {\bf 151B}, 375 (1985) \\
   \> A. Santamaria
and J. W. F. Valle, {\it Phys. Rev. Lett.} {\bf 60}, 397 (1988) \\
   \> A. Masiero
and J. W. F. Valle, {\it Phys. Lett.} {\bf 251B}, 273 (1990) \\
   \> J. Ellis, et. al. {\it Phys. Lett.} {\bf 150B}, 142 (1985) \\
$\left [ 7 \right ]$ \> M. Gell-Mann,
P. Ramond and
R. Slansky, in {\it Supergravity}, ed. F. van Nieuwenhuizen and \\
 \> D. Freedman (North Holland, 1979), p. 315 \\
  \> T. Yanagida, in {\it Proceedings of
the Workshop on Unified Theory and Baryon Number} \\
 \> {\it in the Universe}, ed.
A. Sawada and H. Sugawara, (KEK, Tskuba, Japan, 1979) \\
 \> R. N. Mohapatra
and G. Senjanovic, {\it Phys. Rev. Lett.} {\bf 44}, 912 (1980)\\
$\left [ 8 \right ]$ \> J. C. Pati and A. Salam, {\it Phys. Rev.} {\bf D 10},
275 (1974) \\
 \> R. N. Mohapatra and J. C. Pati,
{\it Phys. Rev.} {\bf D 11}, 566, 2558 (1975) \\
 \> G. Senjanovic and
R. N. Mohapatra, {\it Phys. Rev.} {\bf D 12}, 1502 (1975) \\
$\left [ 9 \right ]$ \> R. N. Mohapatra
and G. Senjanovic, {\it Phys. Rev. Lett.} {\bf 44}, 912 (1980);\\
 \> $~~~~~~~~~~~~~~~~~~~~~~~~~~~~~~~~~~~~~~~~~~~~~~~~~~~~~~${\it Phys. Rev.}
{\bf D 23} 165 (1981) \\
$\left [ 10 \right ]$ \> M. Cvetic
and J. C. Pati, {\it Phys. Lett.} 135B, 57 (1984) \\
$\left [ 11 \right ]$ \> R. N. Mohapatra, {\it Prog. in Part. and Fields}
{\bf 26}, 1 (1991) \\
 \> R. M. Francis,
M. Frank and C. S. Kalman, {\it Phys. Rev.} {\bf D 43}, 2369 (1991) \\
   \> Y. J. Ahn, {\it Phys. Lett.} {\bf 149B}, 337 (1984)\\
$\left [ 12 \right ]$ \> M. Cvetic, {\it Phys. Lett.} {\bf 164B}, 55 (1985) \\
$\left [ 13 \right ]$ \> D. Chang, R.N. Mohapatra and M.K. Parida {\it Phys.
Rev. Lett.} {\bf 52} 1072 (1984)\\
$\left [ 14 \right ]$ \> S. Coleman and
E. Weinberg, {\it Phys. Rev.} {\bf D 7}, 1888 (1973) \\
$\left [ 15 \right ]$ \> J. Ellis,
G. Ridolfi and F. Zwirner, {\it Phys. Lett.} {\bf 257B}, 83 (1991) \\
$\left [ 16 \right ]$ \> U. Ellwanger
and M. Rausch de Traubenberg, {\it Z. Phys.} {\bf C 53}, 521 (1992)
\end{tabbing}
\end{document}